\documentclass[sigconf]{acmart}

\usepackage{booktabs} 
\usepackage{graphicx}

\usepackage{algorithm}
\usepackage{algorithmic}
\usepackage{mathrsfs}

\newfont{\affaddr}{phvr8t at 10pt}

\newenvironment{myitemize2}[1][]{
\begin{list}{$\bullet$}
    {
     \setlength{\leftmargin}{5mm}     
     \setlength{\parsep}{0.5mm}         
     \setlength{\topsep}{0mm}         
     \setlength{\itemsep}{0mm}        
     \setlength{\labelsep}{0.5em}     
     \setlength{\itemindent}{0mm}    
     \setlength{\listparindent}{6mm} 
    }}
{\end{list}}

\copyrightyear{2017} 
\acmYear{2017} 
\setcopyright{acmcopyright}
\acmConference{CIKM'17 }{November 6--10, 2017}{Singapore, Singapore}\acmPrice{15.00}\acmDOI{10.1145/3132847.3133163}
\acmISBN{978-1-4503-4918-5/17/11}
\begin{document}
\title{Session-aware Information Embedding for E-commerce Product Recommendation}
\author{
%
%
{Chen Wu$^{1}$, Ming Yan$^{1}$, Luo Si$^{1}$}\\
      \affaddr{$^{1}$Alibaba Group, 969 West Wenyi Road, Hangzhou 311121, China}\\
       {\{wuchen.wc, ym119608, luo.si\}@alibaba-inc.com}\\
}

\begin{abstract}
Most of the existing recommender systems assume that user's visiting history can be constantly recorded. However, in recent online services, the user identification may be usually unknown and only limited online user behaviors can be used. It is of great importance to model the temporal online user behaviors and conduct recommendation for the anonymous users. In this paper, we propose a list-wise deep neural network based architecture to model the limited user behaviors within each session. To train the model efficiently, we first design a session embedding method to pre-train a session representation, which incorporates different kinds of user search behaviors such as clicks and views. Based on the learnt session representation, we further propose a list-wise ranking model to generate the recommendation result for each anonymous user session. We conduct quantitative experiments on a recently published dataset from an e-commerce company. The evaluation results validate the effectiveness of the proposed method, which can outperform the state-of-the-art.
\end{abstract}

%
%



\maketitle

\vspace{-1mm}
\section{Introduction}

Nowadays, hundreds of millions of people use the e-commerce website for satisfying their daily demands. Due to some privacy reason, more and more users now prefer to surf the e-commerce website anonymously without login in. For example, in the released dataset from real search engine of DIGNETIA~\footnote{\small{https://competition.codalab.org/competitions/11161}}, about 62.3\% of all user sessions are non-logged users. For these users, there exist no historical behavior records  for them and it is difficult for the recommender system to provide accurate recommendation due to a lack of adequate user behaviors.

However, to find the right product, users in e-commerce website usually leave valuable online footprints when searching for it. For example, they may trigger multiple queries within each session or conversation to find a better result and click different products of the same category back and forth to compare the prices. Although limited, these instant search behaviors provide valuable information to understand the user's search intention, which can also help the system to better adapt to the user's subsequent information needs. Therefore, the aim of this paper is to generate recommendations for the anonymous users in e-commerce website, by leveraging the user's temporal behavioral information within the current session, such as clicks, views and purchases.

Traditional recommender systems mainly rely on the long-term user behavior history for recommendation, which follow the idea of collaborative filtering~\cite{su2009survey}. However, the instant online user-item interactions are not well considered, especially in the e-commerce search scenario. The challenge is two-fold: (1) the available user online information is sparse and limited; (2) different kinds of user behaviors exist in the e-commerce website, such as clicks, views and purchases, how to combine all the different user behaviors for a better user understanding.

To address the above challenges, we propose a list-wise deep learning model for solution. First, we present a session embedding method to represent each session with all the available user online behaviors. Different kinds of user behaviors are separately embedded and combined together with multi-layer deep neural networks.
Based on the learnt session representation, a dependent list-wise ranking model is then used to calculate the relevance between the session and the candidate products. To accelerate the training process, the pre-train mechanism is introduced in list-wise ranking. 

The main contributions of this paper can be summarized as: 1) we propose to address the cold-start recommendation problem for the anonymous users in e-commerce scenario by leveraging the diverse online search behaviors within each session; 2) a list-wise deep neural network based framework with pre-training mechanism is designed to model the user's online session behaviors for recommendation; 3) quantitative experiments on a real-world dataset demonstrate the effectiveness of the proposed method over the state-of-the-art recommendation methods.


\vspace{-3mm}
\section{Related Work}
Research on recommender systems has been studied for years, in which collaborative filtering is one of the most popular techniques. It relies on the users' collaborative behaviors on items for recommendation, among which KNN~\cite{sarwar2001item} and latent factor models~\cite{koren2008factorization} are the traditional state-of-the-art methods. To further base the recommendation on the implicit feedback, a generic learning algorithm BPR~\cite{rendle2009bpr} is proposed and applied to the KNN and MF methods. Recently, deep learning has also been applied to the recommender systems. For example, \cite{cheng2016wide} presents a wide and deep learning method to jointly combine the id features and continuous features. 


Traditional collaborative filtering-based methods usually face the problem of data sparsity and will fail for the cold-start users. The proposed solution aims to leverage the instant online search behaviors within session for recommendation, which helps to address the new user problem. Some recent research also attempts to model the user's sequential session behaviors. In \cite{hidasi2015session}, a RNN-based method is proposed to model the sequential session behaviors for more accurate recommendation. \cite{yu2016dynamic} provides a dynamic recurrent model for next basket recommendation. These methods mainly generate recommendations based on a sequence of explicit behavior records, but not distinguish among different types of behaviors. Our proposed method formulate this task as a learning to rank problem, and different types of session information are combined in a list-wise ranking framework. 

\vspace{-1mm}
\section{SESSION-AWARE RECOMMENDATION}


\vspace{1mm}
\hspace{-0.5mm}\textbf{DEFINITION 1} (\textsc{Session}): \emph{We define user session as an uninterrupted sequence of user activities (e.g. clicks, purchases) in the e-commerce system, which is common in the web search field. }

Given a collection of anonymous user session logs in e-commerce website ( i.e., user clicks, views and purchases), and a collection of presented products, the goal of the session-aware recommendation is to re-rank the presented products according to the user's previous online behaviors within the same session.

\vspace{-3mm}
\subsection{Deep ListNet Ranking Framework}

The overall solution framework mainly consists of two parts: \emph{Session Information Embedding (S-IE)} and \emph{List-wise Ranking with S-IE}. In the first part, different types of session behaviors are combined to obtain a compact representation for each session. The main reason behind this part lies in that: 1) to obtain an accurate session embedding in a supervised way, which can help give a better initialization for the subsequent ranking model; 2) to provide a session embedding and item embeddings, which serves as a pre-training mechanism to accelerate the ranking procedure in the next part. To improve the recommendation performance, the second part is designed to further re-rank the presented products under a multi-layer list-wise ranking framework. 

\vspace{-3mm}
\subsection{Session Information Embedding (S-IE)}

\begin{figure}
\centering
\includegraphics[width=0.45\textwidth]{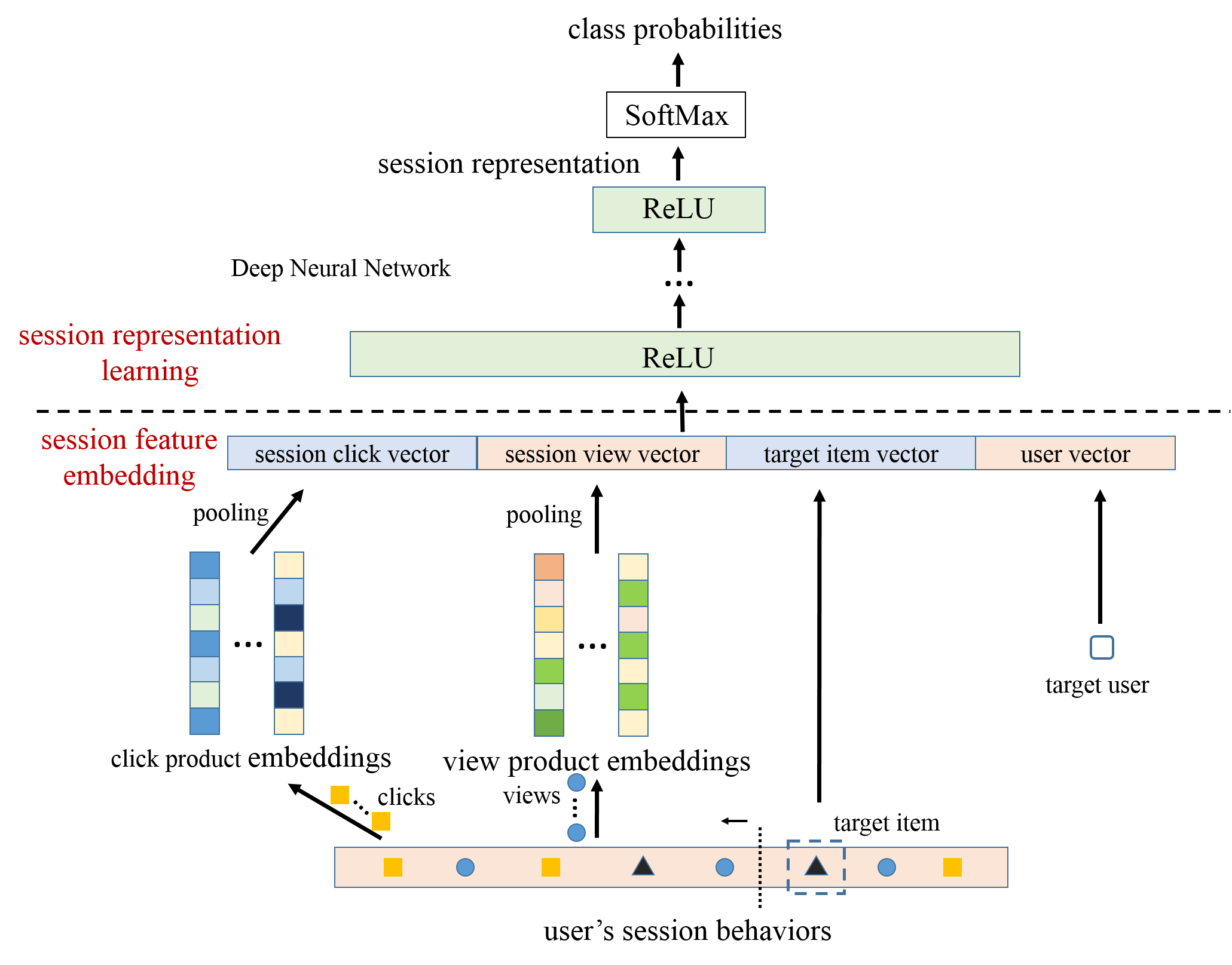}\vspace{-2mm}
\caption{session information embedding model.}\vspace{-3mm}
\label{fig:sie}
\end{figure}

This section introduces how we combine the different types of behaviors within each session to obtain a compact session representation. The overview of this model is illustrated in Figure~\ref{fig:sie}, which mainly consists of two parts: session feature embedding and supervised session representation learning. 

In session feature embedding, given a target user $u$ and a presented item $i$ in session $s$, it is reasonable to assume that user $u$'s click and view behaviors before this item presentation within the session are very significant. Therefore, we design an embedding method to take advantage of all the available session behaviors with behavior vectors. Specifically, we aggregate the user's previous clicks and views separately, and use a pooling strategy (max pooling or average pooling) to generate a session click vector and view vector. Then concatenated with the user id embedding and item id embedding, the final embedding vector can well capture the user intention in this session.

To better combine these different types of user behaviors for a comprehensive session understanding, a deep neural network method with full-connected relu layers is further used for learning combination weights on the derived behavior vectors. We supervise the learning of session representation as a two-class classification problem and the real user's click and purchase behaviors are used as supervision. Specifically, we use the total click behaviors and resample the purchase behaviors three times as the positive samples\footnote{\small{Resampling the purchase behaviors shows good performance in practice.}}. The presentation behaviors with no clicks are used as negative samples and negative sampling is also used for efficiency consideration. We use the cross-entropy loss as our objective function and the network parameters are updated accordingly.

Finally, we extract the last relu layer representation as the session representation. For each session $s$, after the session representation learning, we obtain its completed session representation $\textbf{s}$. Actually, the class probabilities from the softmax layer can already give a coarse ranking for the candidate products in each session. 

\vspace{-2mm}
\subsection{List-wise Ranking with S-IE}

\begin{figure}
\centering
\includegraphics[width=0.4\textwidth]{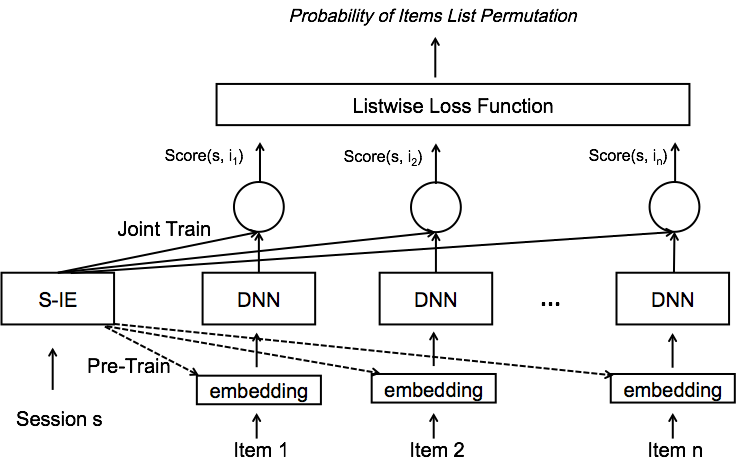}\vspace{-2mm}
\caption{Deep ListNet with SIE}\vspace{-3mm}
\label{fig:frame}
\end{figure}

In session information embedding stage, the obtained coarse ranking does not consider the relative item order within each user session. Therefore, to further refine the recommendation result, we propose a list-wise dependent ranking method to calculate the relevance score between the obtained session representation and the candidate items. To make our list-wise ranking method more computationally efficient and take the most advantage of the obtained session representation, we depend our ranking model on the session representation from the S-IE part and fine-tune the top item embeddings in a list-wise ranking framework. The overview of the ranking framework is presented in Figure~\ref{fig:frame}.

The main goal of this part is actually to match the obtained session representation with representations of the candidate items. The idea is analogous to that of item retrieval, only that there is no query information in our recommendation scenario and we rely on the derived session representation for matching. The proposed list-wise learning method is based on the typical ListNet method~\cite{cao2007learning}, and a full-connected DNN network is introduced to fine-tune the item representations with a list-wise ranking loss. 

Specifically, given a user session $s$ and a list of presented items $I=\{i_1, ..., i_n\}$, we first pass the items $I$ to an embedding layer, so as to obtain a fixed-length vector representation of the items. To better couple with the session representation from the first part, we also pre-train the item embeddings with the SIE model and use the embeddings as the initial item representations. Then, to make it possible to match the session representation $\textbf{s}$ with the item representation $\textbf{i}$ in the same latent space, a full-connected DNN layer is further introduced to project the item embeddings to the session space, which share the same dimension size with the session representation. The score function for a typical session $s$ and item $i$ can be given directly as: $score(s, i)=\textbf{s}\cdot\textbf{i}$.

For the list-wise ranking, we denote the ranking function based on the deep neural network model $w$ as $f_w$. Given a training sample $x^{i}_j$ for item $j$ of session $i$, $f_\omega(x^{i}_j)=score(i, j)$ assigns a score for each sample. For the total item list of session $i$, the ranking function  $f_\omega$ can generate a score list $z^{i}(f_\omega)=(f_\omega(x^{i}_1), f_\omega(x^{i}_2),..., f_\omega(x^{i}_n))$.  

In e-commerce recommendation, people usually just search for the first several pages for their needs and the top-ranked results are what actually matters much. Therefore, we use the typical top-k probability model to calculate the list-wise loss function, which only considers the list order of the top-ranked results. Ranked by the scores,  the top k probability of items is calculated as
\begin{equation}\label{equ:1}
P_{z^{i}(f_\omega)}(\mathbb{G}_k(j_1,j_2,...,j_k))=\prod^{k}_{t=1}\frac{exp(f_\omega(x^{i}_{j_i}))}{\sum^{n^{i}}_{l=t}(exp(f_\omega(x^{i}_{j_i}))}\qquad
\end{equation}
where $\mathbb{G}_k(j_1,j_2,...,j_k)$ is the top k sub-group and $n^i$ is the total sample number of session $i$.

With Cross Entropy as metric, the loss for each session list becomes:
\begin{equation}\label{equ:2}
L(y^{i},z^{i}(f_\omega))=-\sum_{\forall{g}\in{\mathbb{G}_k}}P_{y^{(i)}}(g)log(P_{z^(i)(f_\omega)}(g))
\end{equation}
where $y^i$ is the actual behavior label, and $y^i_j\in\{0,1,2\}$ represent the presentation, click and purchase behavior, respectively.

\begin{algorithm}[h]
\caption{Learning Algorithm of ListNet-SIE}
\begin{algorithmic}[1]
\REQUIRE ~~\\ 
Training data:\{($x^{1}$, $y^{1}$),($x^{2}$, $y^{2}$),...,($x^{n}$, $y^{n}$)\}\\
Parameters: number of iterations T and learning rate $\eta$\\
Initialize parameter $\omega$
\FOR{$t=1$ to $T$}
\FOR{$i=1$ to $n$}
\STATE Input $x^{i}$ joint S-IE $s^{i}$ to Neural Network and compute score list $z^{i}(f_\omega)$ with current $\omega$
\STATE Compute gradient $\Delta\omega$ using Eq. ( ~\ref{equ:3} )
\STATE Update $\omega=\omega-\eta\times\Delta\omega$
\ENDFOR
\ENDFOR
\ENSURE Neural Network model $\omega$\\ 
\end{algorithmic}
\end{algorithm}  

The gradient of the loss function with respect to parameter $\omega$ is
\begin{equation}\label{equ:3}
\Delta\omega=\frac{\partial{L(y^{i},z^{i}(f_\omega))}}{\partial\omega}=-\sum_{\forall{g}\in{\mathbb{G}_k}}\frac{\partial{P_{z^(i)(f_\omega)}(g)}(f_\omega))}{\partial\omega}\frac{P_{y^{(i)}}(g)}{P_{z^(i)(f_\omega)}(g)}\qquad
\end{equation} 
Each session with the contained item behaviors is treated as a list-wise training sample and the Gradient Descent method can be then used to update the model parameters $\omega$ accordingly. Algorithm 1 shows the learning algorithm of ListNet-SIE. Finally, with the learnt DNN weight $W^T_{DNN}$ and the item embedding $\textbf{i}_e$, the ranking score for the target item $i$ in session $s$ is: 
\begin{equation}\label{equ:3}
Score(s, i)=\textbf{s}\cdot W^T_{DNN}\textbf{i}_e 
\end{equation}
The presented results are finally re-ranked according to $score(s, i)$.

\section{Experiments}
\subsection{Dataset}
For the task of e-commerce product recommendation, we perform experiments on a publicly released dataset of the track 2 of CIKM Cup 2016~\footnote{\small{https://competitions.codalab.org/competitions/11161}}. This dataset contains anonymized users' six-month click, view and purchase records on an e-commerce search engine, and the user session is extracted using a 1-hour of inactivity heuristic. To focus on our session-aware recommendation scenario, we selected all the ``query-less" sessions in this dataset and filtered out the sessions with less than two queries. The last query in each user session is reserved as the test set, the others are used for training. The statistics of the dataset is summarized in Table~\ref{tab:stat1}.

\vspace{-1mm}
\subsection{Experimental Settings}
In session information embedding, the embedding size of each type of behavior is 200. Three relu layers with hidden unit size of 800, 200, 100 are used to get good performance. In supervised learning, negative sampling is adopted and the ratio of positive v.s. negative samples is 1:5, which can already give an optimal result. In list-wise ranking, we use two full-connected neural network layers with sigmoid activation function, where the hidden unit sizes are both 100. For the top k probability model, we choose k as 10 for the page presentation situation and efficiency consideration. The learning rate $\eta$ is set to 0.001 to ensure the convergence of the algorithm.

\begin{table}[t] \small
\centering
\caption{\label{tab:stat1} Basic statistics of the dataset.} \vspace{-3mm}
\begin{tabular}{c|c}
\hline
    Statistics & Value \\
\hline
    \#users & 21,930\\
    \#sessions & 27,644\\
    \#query-less queries & 82,368\\
    \#presented products & 112,685 \\
    \#click logs  & 160,521\\
    \#view logs & 181,180\\
    \#purchase records & 43,666\\
    \#avg.(show items) per query & 146.5\\
\hline
\end{tabular} \vspace{-3mm}
\end{table} 

To evaluate the effectiveness of the proposed two-stage solution, we implemented five typical baselines and two different settings of our solution. The seven examined methods are listed as below:
\begin{myitemize2}
  \item \emph{Popularity}: the simple personalized re-ranking baseline provided by the official site;
  \item \emph{KNN}: the typical item-based collaborative filtering recommendation algorithm~\cite{karypis2001evaluation};
  \item \emph{LFM}: state-of-the-art Latent Factor Model \cite{koren2008factorization}, which is mainly designed to address the sparsity problem;
  \item \emph{BPR}: the typical bayesian personalized ranking method with the implicit feedback~\cite{rendle2009bpr};
  \item \emph{S-RNN}: a typical session-based recommendation method with recurrent neural networks~\cite{hidasi2015session};
  \item \emph{SIE}: the proposed solution that directly use the class probabilities from session information embedding part for ranking;
  \item \emph{ListRank}: the proposed solution considers both the session information embedding and deep ListNet ranking.
\end{myitemize2}

We adopt the NDCG metrics, which also serves as the final competition goal. The NDCG score is defined on each test query, and the final result is averaged over all the test queries.  

\vspace{-1mm}
\subsection{Experimental Results and Analysis}
\begin{figure}
\begin{center}
\includegraphics[width=0.38\textwidth]{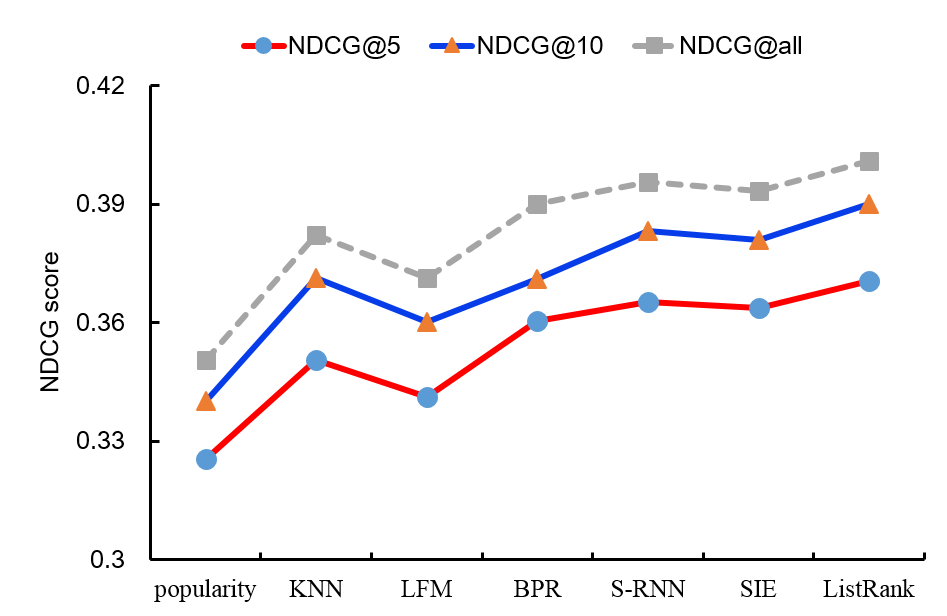}
\end{center}
\vspace{-3.5mm}
\caption{The NDCG results of all the examined methods.}  \vspace{-5mm}
\label{fig:res1}
\end{figure}

The NDCG results of all the examined methods are shown in Fig.~\ref{fig:res1}. Thus it can be seen: (1) compared with other typical methods, LFM and KNN get the worst performance, which may be due to that the user behaviors are not adequate in our session-based scenario; (2) by considering the online session information, \emph{SIE} and our proposed methods can outperform other baselines, while the proposed \emph{ListRank} method is slightly better for the consideration of combination of different types of session behaviors; (3) \emph{ListRank} further improves the performance over the \emph{SIE} method, which validates the effectiveness of the list-wise ranking procedure.

\textbf{Influence of different behavior embeddings.}  To further investigate what session information contributes most to the final performance,  we also examine with different session vector representations in the SIE part: (1) with no previous session information; (2) with only previous click vector representation; (3) with only previous view vector representation; and (4) with both representations.  The comparative results are shown in Table~\ref{tab:bev}.

\begin{table}[t] \small
\centering \caption{The NDCG@all results with different behavior vector representations in SIE part.} \vspace{-3mm}
{
\begin{tabular}{|c|c|c|c|c|}
  \hline
   & no click \& view & only click & only view & both \\
  \hline
  SIE & 0.359 & 0.385 & 0.376 & \textbf{0.393} \\
  \hline
   ListRank & 0.368 & 0.393 & 0.384 & \textbf{0.400} \\
  \hline
\end{tabular}
}
\vspace{-3mm}
\label{tab:bev}
\end{table}

From the results, we can see that: (1) with either click or view information embedded in session representation, the recommendation performance gets improved, which shows the importance of the previous session behaviors. (2) The click behavior is slightly more important than the view behavior, which may be due to that only click and purchase behaviors indicate the positive labels to calculate the NDCG formula. (3) Combining both the click and view behaviors for session representation learning leads to better recommendation performance in both proposed methods, which also shows the effectiveness of our method for combining different types of online session behaviors.

\section{Conclusions}
In this paper, we have proposed a novel session-aware recommendation method to address the recommendation problem of anonymous users surfing in e-commerce website. The overall framework mainly consists of two parts: session information embedding and list-wise ranking. Firstly we embed each session with multiple user session behaviors and different types of online session behaviors are combined for a comprehensive session understanding. Then a deep list-wise ranking model is used to re-rank the presented results for better recommendation. The experiment results validate the effectiveness of the proposed method in modeling different types of online session information. In the future, we are also considering designing dynamic neural network models such as LSTM to better understand the user's sequential online behaviors.

%
\bibliographystyle{abbrv}
\bibliography{sigproc}  
%
%

\end{document}